\documentclass[12pt,preprint]{aastex}
\shorttitle{Stellar Oscillations in $\beta$ Gem}
\shortauthors{Hatzes and Zechmeister}
\begin{document}

\title{The Discovery of Stellar Oscillations in the Planet Hosting Giant Star
$\beta$~Geminorum}

\author{Artie P. Hatzes}
\and
\author{Mathias Zechmeister}
\affil{Th\"uringer Landessternwarte, D - 07778 Tautenburg, Germany}
\email{artie@tls-tautenburg.de}
\email{zechmeister@tls-tautenburg.de}

\begin{abstract}
We present the results of a long time series of precise stellar radial velocity measurements  of
the planet hosting K giant star $\beta$~Geminorum. A total of 20 hours 
of observations
spanning three  nights were obtained and the radial velocity variations show the presence of solar-like
stellar oscillations. Our period analysis yields six  
significant pulsation modes that have frequencies in the 
range of 30 -- 150 $\mu$Hz. The  dominant mode is at a frequency of 86.9 $\mu$Hz
and has an amplitude of 5.3 m\,s$^{-1}$. These values are consistent with 
stellar oscillations for a giant star with a stellar mass  of 
$\approx$ 2 $M_\odot$.
This stellar mass implies a companion minimum mass of 2.6 $M_{Jupiter}$.
$\beta$~Gem is the first planet hosting giant star 
in which multi-periodic stellar oscillations have been detected.
The study of stellar oscillations in planet hosting giant stars may provide
an independent, and more accurate  determination of the stellar mass.
\end{abstract}


\section{Introduction}
\label{intro}

Radial Velocity (RV) surveys have discovered a number of giant planets in orbit
around giant stars (Frink et al. 2002; Sato et al. 2003, Setiawan et al. 2003,
Hatzes et al. 2005). These discoveries are important because the
evolved host stars often have masses in the range 1--3 $M_\odot$.  Precise stellar 
radial velocity measurements of 
main sequence stars in this mass range are difficult due to the high effective 
temperatures of the photosphere
and the paucity of stellar lines. Furthermore, these lines are often broadened by
high rates of stellar rotation. Consequently, most 
RV surveys have focused primarily on stars later than about spectral type F6.
The study of planets around giant stars
can thus give us valuable clues as to the process or planet formation around
stars more massive than the sun, {\it if} one can determine an accurate stellar mass.
For giant stars  this can be  difficult.
Evolutionary tracks 
of main sequence stars spanning a 
wide range of masses all converge to the giant branch in 
the color magnitude diagram.  One has to rely on stellar
evolutionary tracks
which are model dependent, and these in turn rely on accurate determinations
of such stellar parameters as effective temperature, surface gravity, 
abundance, and absolute luminosity. 

The mass determination of Arcturus offers us a good example.
Using spectral analysis M\"ackle et al. (1975) determined a stellar mass
of 0.1 -- 0.6 $M_\odot$. The analysis of Martin (1977) yielded a mass in the 
range 0.6 -- 1.3 $M_\odot$. This is consistent with a later mass determination
of 0.95 $M_\odot$ by Bell, Edvardsson, and Gustafson (1985). These authors
noted that due to uncertainties in the surface gravity the 
stellar mass could be as low as
0.7 $M_\odot$. An accurate stellar mass is not only important for comparing
results to planet formation theories, but it is also required to calculate
the companion mass. Planets around intermediate mass
stars  have masses in the range 3--10 $M_{Jupiter}$ and many lie on the deuterium
burning border which separates brown dwarfs from planets (approximately 13 $M_{Jupiter}$). 
More accurate determinations of the stellar mass may establish if a
companion has a mass that is still consistent with a bona fide planet
or is  certainly a brown dwarf, even if the orbital inclination were to be nearly
90$^\circ$ (i.e sin $i$ = 90$^\circ$).

One of the best ways of determining stellar mass, outside of dynamical methods, is
via asteroseismology. This is also the most accurate method
for determining the masses of isolated stars.
The stellar oscillations can be used to
derive such 
fundamental parameters as the stellar mass, radius, age, and, 
depending on the number of modes detected, the internal structure.
Asteroseismology has been used with spectacular success on white dwarf stars
using multi-site photometric campaigns
(e.g. Castanheira et al. 2004).
More recently, thanks to an increase in the precision of stellar RV
measurements, asteroseismology has been applied with some success
to solar-like stars (e.g. Bedding et al. 2006; Bazot et al. 2005).

It is well established that many cool giants exhibit short period
RV or photometric 
variations with periods ranging from hours (e.g. Hatzes \& Cochran 1994b;
Frandsen et al. 2002; de Ridder et al. 2006) to days 
(e.g. Hatzes \& Cochran 1994a, Retter et al. 2003).
These  periods are consistent with p-mode oscillations in giant stars.
Although many of these modes have not been identified with certainty, the observed
periods seem to be consistent with radial fundamental or overtone modes 
(Hatzes \& Cochran 1994a), 
although nonradial modes can still not be excluded.
The fact that extrasolar planets have been discovered around a class of 
stars known to exhibit stellar oscillations opens up the exciting 
possibility of using these stellar oscillations as an independent means
of deriving important properties of the planet host star.

Long period variations with a period of 545 days were discovered in 
$\beta$~Gem by Hatzes \& Cochran (1993). One proposed explanation was that
these were due to a planet with a minimum mass of 2.9 $M_{Jupiter}$,
assuming a stellar mass of 2.8 $M_\odot$.
Over a decade later Hatzes et al. (2006) and Reffert et al. (2006) 
confirmed that these variations were in fact due to a planet in orbit with
a revised period of 590 days. Hatzes et al. (2006) used a more recent
stellar mass  determination of 1.7 $M_\odot$ to derive a companion 
mass of 2.3 $M_{Jupiter}$.  However, given the difficulties of deriving 
stellar masses from evolutionary tracks of giants, $\beta$~Gem 
could easily have a much lower or higher mass. If one is interested in understanding the stellar mass
dependence of planet formation it is important to know if indeed $\beta$~Gem has a mass
in the intermediate ($\approx$ 2 $M_\odot$) range.

Here we present  time series RV measurements for $\beta$~Gem taken on three
nights. The star shows RV variability consistent with stellar 
oscillations. We then use these oscillation frequencies to estimate the stellar
mass.

\section{Observations}

	 A long time series of spectral observations were made of
$\beta$~Gem using the coude echelle spectrograph of the 2m Alfred Jensch
Telescope of the Th\"uringer Landessternwarte (Thuringia State Observatory).
Precise stellar radial velocity measurements were achieved by using an 
iodine absorption cell placed in the optical light path to provide the
wavelength reference. A detailed description of the instrumental setup 
and data reduction and analysis process can be found in Hatzes et al. (2005).

	Exposure times were 90 secs and with a CCD readout time of
70 secs. The signal-to-noise ratio ($S/N$) of the observations depended on atmospheric
transparency and seeing conditions. The $S/N$ for our observations
ranged from about 100 to 300 per pixel.  
A total of over 20 hours of observations were made
$\beta$~Gem spanning 3 different nights. 
Table 1 gives the journal of observations which includes the
Julian day at the start of the time series,  the length of the time
series, and the number of spectra that were obtained. The last column
is the nightly rms scatter of the 
RV measurements about a mult-sine fit to the data (see below).

	Figure~\ref{f1} shows the time series of the RV measurements on
the 3 nights. It is evident from the figure that $\beta$~Gem shows
low amplitude, periodic variability on short time scales. It is also clear
that a single period  cannot reproduce the observed variations.

\section{Period Analysis}
	
	A period analysis was performed on the full data set using the program {\it Period04}
(Lenz \& Breger 2004).
This program offers convenient means of  searching for multiple periods
in data via a pre-whitening  procedure. A 
sine wave fit is made to the data using the
dominant period found by Fourier analysis. This is subtracted and 
additional periods are found in the residuals by further Fourier analysis. 
Finally, {\it Period04} can be used to improve the solution by performing 
a simultaneous least squares fit using a sum
of sine functions with the initial guess periods found in the
pre-whitening procedure. 

The Fourier analysis revealed a long period component corresponding to a period of
3 days. Our data string is too short to establish if this is real or possibly an alias
of a shorter period. Since we are searching for periods comparable to, or shorter than our 
nightly coverage this long-period component was subtracted at the start of the 
period analysis.

\begin{figure}
\plotone{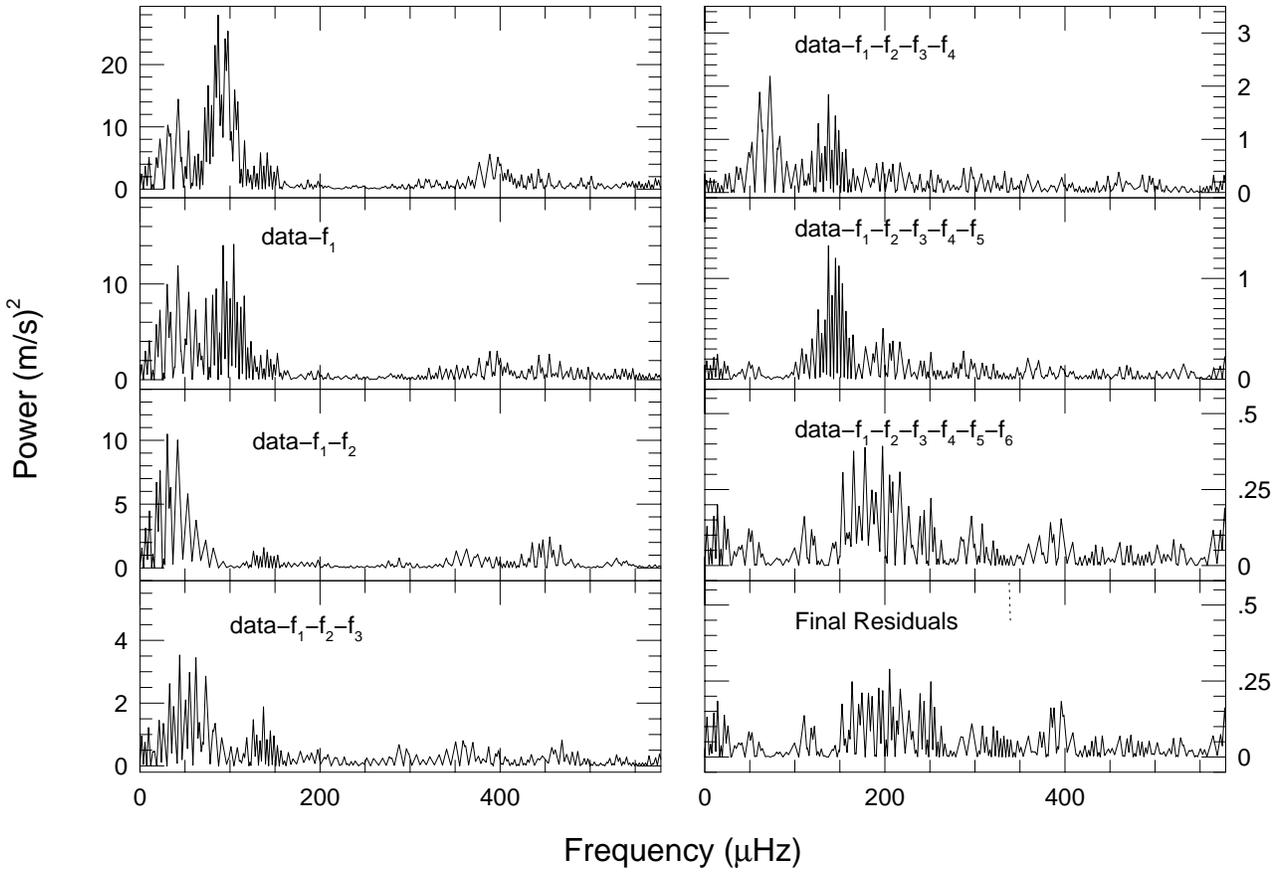}
\figcaption[]{Power spectra of the RV data at each step of the
pre-whitening procedure. The top left panel is for the raw RV data.
Each successive lower panel shows the power spectrum
after subtracting the contribution of the dominant mode
from the previous (upper adjacent) panel. The lower right panel is for the final
RV residuals after removal of the possible
seventh mode ($f_7$).  Note the change in y-axis scale for each panel.
\label{f2}}
\end{figure}

        Figure~\ref{f2} shows the power spectra derived using the discrete
Fourier transform at
each step of the pre-whitening procedure.
The top panel is for the
``raw'' time series, and each subsequent lower panel is the power spectrum
after removal of the dominant frequency.

The statistical significance of the periods were assessed using a
``bootstrap randomization technique''.
The RV values was  randomly shuffled
keeping the times fixed and a Scargle periodogram calculated
(Scargle 1982). A Scargle periodogram was used since the power is a measure
of the statistical significance of a signal. After a large
number of shuffles (200,000) the fraction of random periodograms
having Scargle power greater than the data periodogram gave an
estimate of the false alarm probability (FAP)
that a signal is due purely to noise.
Our bootstrap analysis indicates that a Scargle power, $P$ =  15 corresponds
to a FAP = 10$^{-5}$. 
The pre-whitening procedure continued until
the Fourier analysis
found a period which we deemed not to be significant. For this analysis we
considered any signal having a   FAP $>$ 10$^{-4}$ as not significant.
The dominant peak in each panel of
Figure~\ref{f2} had Scargle power greater than 15, with the exception of the
last ($f_7$) mode.

	Table 2 lists all the periods, frequencies, and corresponding amplitudes
for all the statistically significant periods found by our analysis.  The modes are listed in the
order they were found by the pre-whitening procedure. The first six frequencies are highly significant
having FAP $<$ 5 $\times$ 10$^{-6}$. In other words after 2 $\times$ 10$^{5}$
shuffles of the bootstrap there was no instance when the power in the random periodograms
had higher power than the data periodograms. The least significant frequency is the one
at $\nu$ = 193 $\mu$Hz. The false alarm probability of this signal is 1.6  $\times$ 10$^{-3}$ which is
higher than our adopted threshold for significance.
The amplitude of this mode is significantly less than the mean measurement error
(1.5 -- 2 m\,s$^{-1})$.  We thus regard this mode as uncertain and in need of confirmation 
with data spanning a longer time.
The line in Fig.~\ref{f1} shows the multi-component sine fit to the full data set. The rms scatter about the
fit for the individual nights are 1.2, 1.5, and 1.7 m\,s$^{-1}$, respectively.

\section{Stellar Mass Determination}

To estimate the stellar mass we used 
the scaling relations of Kjeldsen \& Bedding (1995) which they showed to
be  valid
for stars covering a wide range of masses and luminosity classes.
In particular 
we will use their expression for the 
frequency of the maximum power:

$$\nu_{max} = { {M/M_\odot} \over {(R/R_\odot)^2 \sqrt{T_{eff}/5777 K} }}
\hskip 5pt 3.05 \hskip 5pt \rm{mHz} $$

We chose to use this expression rather than the one for the
frequency splitting  because of the short time span of our observations.
We are not confident that we have detected all possible modes in $\beta$~Gem
which are necessary for determining a frequency spacing for high order
p-modes.
We take the 87.9 $\mu$Hz frequency as the dominant mode  in the
data. This has the highest amplitude and is the most obvious peak in the periodogram
of the un-whitened data.

The stellar radius of $\beta$~Gem has been measured with long baseline
interferometry. Nordgren, Sudol, \& Mozurkewich (2001) determined an
angular diameter of 7.96 $\pm$ 0.09 mas which corresponds to a radius
of 8.8 $\pm$ 0.1 $R_\odot$ using the Hipparcos distance of 96.74 $\pm$ 0.94
mas.  McWilliam (1990) derived an effective temperature of
4850 K. Using a value of $\nu_{max}$ = 0.0868 mHz results in an stellar mass, $M$ = 2.04 $\pm$ 0.04  $M_\odot$.
The orbital solution for the companion has a mass function, $f(m)$ =
(4.21 $\pm$ 0.48) $\times 10^{-9}$ $M_\odot$ (Hatzes et al. 2006). Using our
nominal stellar mass
results in a minimum mass
for the companion, $m$~sin~$i$ = 2.56 $\pm$ 0.5 $M_{Jupiter}$.

A caveat is in order regarding the uncertainty of our mass determination.
The error is based on the uncertainties
in the stellar parameters. However, the Kjeldsen \& Bedding scaling
relations  are based on assumptions which may not be valid for a giant
star like $\beta$ Gem. Furthermore, due to the short time span of our
measurements the frequency of maximum power may actually be in
an adjacent mode. If the frequency of maximum power is actually in
the modes at $\nu$ = 86.9 $\mu$Hz or $\nu$ = 104.4 $\mu$Hz then
the stellar mass can be as low as 1.85 $M_\odot$ or as  high as
2.42 $M_\odot$. We thus adopt a value of $\pm$0.3 $M_\odot$ as the error in our
mass determination.

\section{Discussion}

	We have detected, for the first time, stellar oscillations in 
a K giant star known to host an extrasolar giant planet. Our analysis
reveals six significant periods, with the dominant mode at a frequency
of 86.8 $\mu$Hz. Using the scaling relations of Kjeldsen \& Bedding (1995)
as well as the interferometric  stellar radius determination results in 
a stellar mass of 2.04 $M_\odot$. This is in reasonable agreement with the stellar
mass of 1.7 $M_\odot$ used by Hatzes et al. (2006) in deriving a companion
mass of 2.3 $M_{Jupiter}$. This provides independent confirmation
that $\beta$~Gem is indeed an intermediate mass star.

The scaling relations of Kjeldson \& Bedding (1995) can also be used
to predict the velocity amplitude:

$$v_{osc} = {{L/L_\odot} \over {M/M_\odot}} (23.4 \pm 1.4) \hskip 5pt 
{\rm cm\,s^{-1}}$$

The absolute $V$-mag of $\beta$~Gem is 1.08 mag. The effective temperature
corresponds to a bolometric correction for a giant star of $BC$ = $-$0.42 
(Cox 2000) which gives a luminosity, $L$ = 42.8 $L_\odot$. The predicted
pulsational velocity amplitude is thus $v_{osc}$  = 5.0 $\pm$ 0.3 m\,s$^{-1}$,
in excellent agreement with the 5.3 $\pm$ 0.38 m\,s$^{-1}$ of the dominant mode.

	The predicted order of the mode, $n$, can be estimated using 
Equation 10 of Kjeldson \& Bedding (1995). For the stellar parameters of $\beta$
Gem this corresponds to $n$ $\approx$
10. The detected modes are most likely high order radial or non-radial modes.

The mean frequency spacing of the modes in Table 2 is 
approximately $\approx$ 20 $\mu$Hz.
This is considerably
higher than the 7.3 $\mu$Hz large spacing expected given the mass and radius of 
$\beta$~Gem. We believe that we have not detected all possible modes
due to the short time span of our measurements.
Deeming (1975) showed that for modes of unequal amplitudes the modes must be separated by
a frequency of 2.5/T, where $T$ is the time span of the measurements. Our measurements
have $T$ $\approx$ 3 days which corresponds to 
$\delta\nu$ = 8.8  $\mu$Hz. Furthermore, we do not know what the mode lifetimes
are. Continuous measurements spanning
a week or more may be required to derive the full oscillation spectrum of $\beta$~Gem.

Our investigation of $\beta$~Gem shows the potential of using stellar
oscillations to determine the stellar mass of giant stars.
We suspect that stellar oscillations in  K giant stars are ubiquitous. If
that
is the case, then an investigation of the stellar oscillations in planet hosting
giant star can be used to determine a more accurate stellar mass.
We are aware that our mass estimate is only an approximation. More modes 
need to be detected in 
$\beta$~Gem and a proper theoretical modeling
is required to derive a more accurate stellar mass. We are currently analyzing additional
RV measurements for this star to search for additional modes.

As is the case with ground-based asteroseismic studies it is very difficult
to get the requisite telescope time needed detect all possible modes.
Observations using multi-site campaigns  are needed  due
to the one day alias at single sites.
This could be difficult
because not all astronomical facilities are equipped for making precise
stellar RV measurements. Furthermore, even with multiple observing sites,
poor weather conditions at one or more sites can still produce data gaps.
In this respect the CoRoT space telescope (Baglin et al. 2001)
will provide a major breakthrough in the study of K giant oscillations.
CoRoT is a 27cm telescope with a dual mission -- asteroseismology on bright
stars and a search for transiting exoplanets in a field of
up to 12,000 stars in the visual magnitude range $V$ = 11--16.
CoRoT can achieve a photometric precision of $\approx$ 2 $\times$
10$^{-4}$ in one hour for a star in the exofield with $V$ =
15.4.  CoRoT was launched on 27 December 2007 and as of this writing the 
light curves from the first observed field had not yet been released.

The Kjeldsen \& Bedding (1995) scaling relations yield a predicted
photometric  amplitude of 100 ppm (10$^{-4}$) for a giant star exhibiting
stellar oscillations like $\beta$~Gem, a precision that can be
reached for most stars in the CoRoT exofield, and 
many of these will be giant stars.
The 150-day, uninterrupted observations of a given long run CoRoT 
field should yield the full oscillation spectrum for stars throughout the
giant branch. 
Given that sub-stellar companions may be common around
giant stars a search for companion around pulsating K giants
found by CoRoT should prove
fruitful for planet formation studies.

\hskip 10pt
\centerline{Acknowledgments}

This work was based on  observations made on the 2m Alfred Jensch Telescope
of the Th\"uringer Landessternwarte Tautenburg. 
We acknowledge the the support of grant 50OW0204
from the Deutsches Zentrum f\"ur Luft- und Raumfahrt e.V. (DLR).
This research has made use of the SIMBAD database,
operated at CDS, Strasbourg, France.

\clearpage

\begin{deluxetable}{cccc}
\tablecaption{Journal of Observations}
\tablewidth{0pt}
\tablehead{
\colhead{Start}        & \colhead{Time Coverage}  & \colhead{$N_{Obs}$}  &   \colhead{$\sigma$} \\
\colhead{(Julian Day)} & \colhead{(hours)}        &                      &   \colhead{(m\,s$^{-1}$)} }
\startdata
2450168.308 & 6.88 & 115 & 1.2 \\
2450170.266 & 5.04 &  74 & 1.5 \\
2450171.259 & 8.64 & 152 & 1.7 \\
\enddata
\end{deluxetable}

\begin{deluxetable}{ccrl}
\tablecaption{Oscillation modes for $\beta$~Gem }
\tablewidth{0pt}
\tablehead{
\colhead{Mode} & \colhead{Period}    & \colhead{Frequency}  & \colhead{Amplitude
}    \\
             & \colhead{(hours)}    & \colhead{$\mu$Hz}  &  \colhead{m\,s$^{-1}$
}   }
\startdata
$f_1$ & 3.20 & 86.91 $\pm$ 0.37    & 5.29 $\pm$ 0.38 \\
$f_2$ & 2.66 & 104.40 $\pm$ 0.49   & 4.09 $\pm$ 0.28 \\
$f_3$ & 9.34 & 29.75  $\pm$ 0.54   & 3.61  $\pm$ 0.22 \\
$f_4$ & 6.23 & 48.41 $\pm$ 1.06    & 1.85  $\pm$ 0.18 \\
$f_5$ & 3.85 & 79.47  $\pm$ 0.76   & 2.58  $\pm$ 0.16 \\
$f_6$ & 2.02 & 149.25  $\pm$ 0.63  & 1.25  $\pm$ 0.15 \\
$f_7$ & 1.56 & 193.63 $\pm$ 3.07   & 0.64  $\pm$ 0.13 \\
\enddata
\end{deluxetable}

\clearpage

\begin{figure}
\plotone{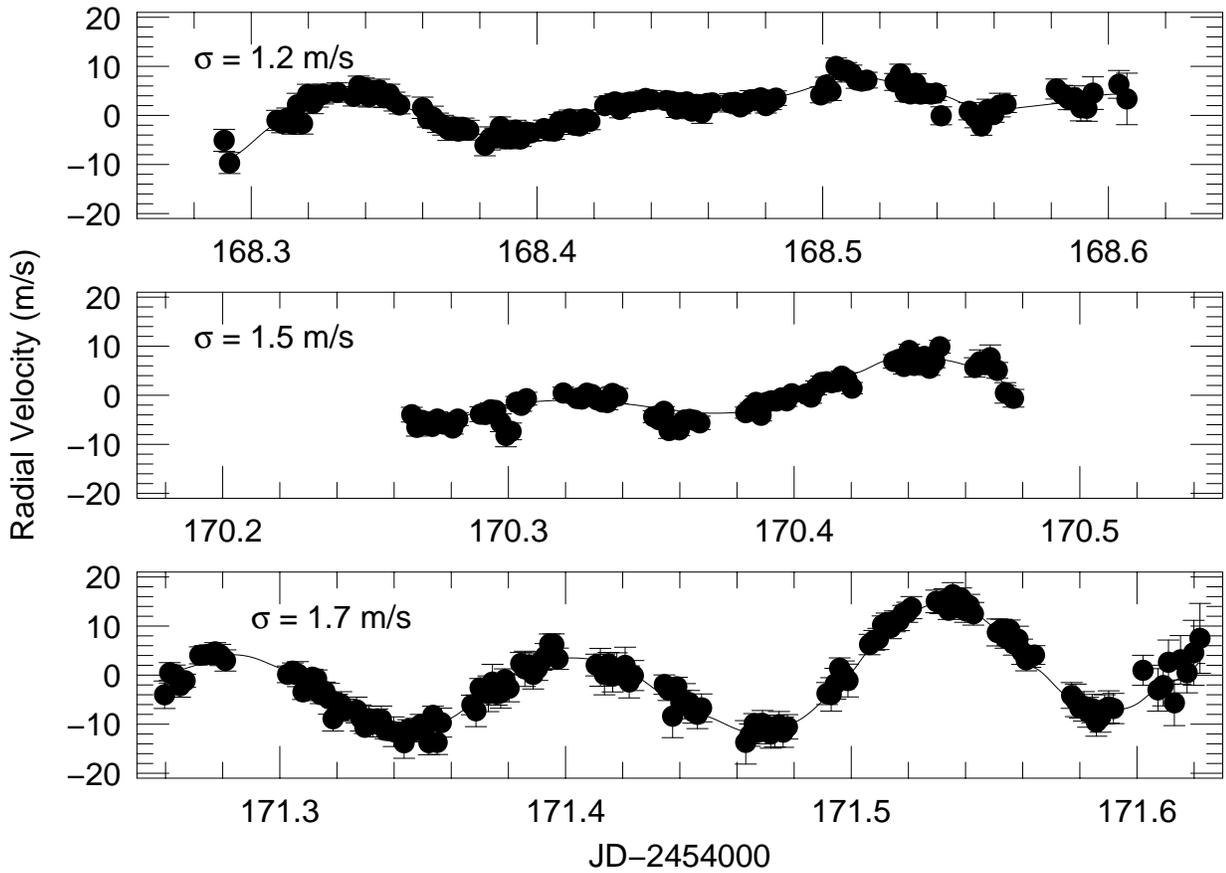}
\figcaption[ftall.ps]{Time series of the RV measurements 
for $\beta$~Gem for three nights. The line represents a 
multi-component sine fit using the six  frequencies of Table 1.
The standard deviations of the RV values about this fit are
1.2, 1.5, and 1.7 m\,s$^{-1}$ on the three nights, respectively.
\label{f1}}
\end{figure}
\clearpage

\end{document}